# Efficient, high-density, carbon-based spinterfaces

F. Djeghloul[1§], G. Garreau[2], M. Gruber[1,3], L. Joly[1], S. Boukari[1], J. Arabski[1], H. Bulou[1], F. Scheurer[1], F. Bertran[4], P. Le Fèvre[4], A. Taleb-Ibrahimi[4], W. Wulfhekel[3], E. Beaurepaire[1], S. Hajjar-Garreau[2], P. Wetzel[2], M. Bowen[1], W. Weber[1*]

[1]IPCMS UMR 7504 CNRS, Université de Strasbourg, 23 rue du Loess, BP 43, 67034 Strasbourg Cedex 2, France

[2]Institut de Science des Matériaux de Mulhouse, CNRS-UMR 7361, Université de Haute-Alsace, 68057 Mulhouse, France

[3]Physikalisches Institut, Karlsruhe Institute of Technology, Wolfgang-Gaede-Strasse 1, 76131 Karlsruhe, Germany

[4]Synchrotron SOLEIL, L'Orme des Merisiers, Saint-Aubin - BP 48, 91192 Gif-sur-Yvette, France

---

[*]Corresponding author. Tel: ++33-3-88107094. E-mail: Wolfgang.Weber@ipcms.unistra.fr (Wolfgang Weber)

**Abstract**

The research field of spintronics has sought, over the past 25 years and through several materials science tracks, a source of highly spin-polarized current at room temperature. Organic spinterfaces, which consist in an interface between a ferromagnetic metal and a molecule, represent the most promising track as demonstrated for a handful of interface candidates. How general is this effect? We deploy topographical and spectroscopic techniques to show that a strongly spin-polarized interface arises already between ferromagnetic cobalt and mere carbon atoms. Scanning tunneling microscopy and spectroscopy show how a dense semiconducting carbon film with a low band gap of about 0.4 eV is formed atop the metallic interface. Spin-resolved photoemission spectroscopy reveals a high degree of spin polarization at room temperature of carbon-induced interface states at the Fermi energy. From both our previous study of cobalt/phthalocyanine spinterfaces and present x-ray photoemission spectroscopy studies of the cobalt/carbon interface, we infer that these highly spin-polarized interface states arise mainly from $sp^2$-bonded carbon atoms. We thus demonstrate the molecule-agnostic, generic nature of the spinterface formation.

**1. Introduction**

The research field of organic spintronics [1] has historically focused on spin-polarized transport across organic semiconductors [2,3], yet over a decade later the mechanism of spin transport invoked to explain experimental results remains elusive, thereby generating controversy [4,5]. Studies of spin-polarized transport across ultrathin dielectric layers within the well-established tunneling regime [6] have remained scarce regarding organic materials [7,8]. This underscores how the counter electrode deposition atop the porous organic layer results in metal interdiffusion that decreases the effective transport distance across the organic layer and can short-circuit the device. Yet such devices could harness quite promising

properties – this time for the spintronics field at large [9] – of interfaces between a ferromagnet and molecules [10-12], which are highlighted by our report of over 80% spin polarization at room temperature when pairing Co with Mn-phthalocyanine (Pc) or $H_2Pc$ molecules [12]. Such interfaces are called organic spinterfaces. This discovery establishes organic spinterfaces as the most promising materials science track toward implementing an ideal spin-polarized current source --- a crucial goal for the field of spintronics over the past 25 years [12]. Yet, since only a handful of ferromagnet/molecule pairs were validated, the question arises how general this effect is.

In the present work, we deploy topographical and spectroscopic techniques to show that a similarly efficient spinterface arises between Co and mere C atoms. We thus demonstrate the molecule-agnostic, generic nature of the spinterface formation. As a perspective, we discuss how the dense semiconducting C film with a low band gap that is formed atop the spinterface could both sidestep the challenge of achieving a structurally sharp top spinterface, and offer opportunities toward nanoscale applications.

## 2. Experimental

The growth of Co onto fcc Cu(001) has been extensively investigated in the past [13-16]. The Cu substrate is first cleaned by several sputtering-cycles and annealing at 800 K. Co films are then deposited from a rod heated by electron beam bombardment. The Cu crystal is maintained at room temperature in order to prevent Cu atoms from segregation during growth, as verified by Low-Energy Ion Scattering (LEIS). With the Cu crystal still at room temperature, carbon is then evaporated from an electron beam evaporator at a deposition rate of 0.02 monolayer (ML)/min. Cobalt and carbon thicknesses are measured by a quartz microbalance and are accurate within 5 % and 15 %, respectively, according to cursory verification using the attenuation of the Co Auger peaks with increasing C coverage as a guideline [12].

Scanning Tunneling Microscopy (STM) and Scanning Tunneling Spectroscopy (STS) (in the Current Imaging Tunneling Spectroscopy (CITS) mode) are performed at room temperature. X-ray Photoemission Spectroscopy (XPS) measurements are carried out using photons from a non-monochromatized Mg K$\alpha$ source (1253.6 eV). The C1s peak is recorded in normal-emission geometry using the same parameters (photon flux, pass energy, etc.) for all samples. The acquisition time is limited to 45 minutes in order to avoid significant contamination under irradiation. Decomposition of the spectra into different components is performed with Gaussian-Lorentzian shaped peaks using XPS Casa software after having subtracted a Shirley-type background. LEIS measurements are carried out with 1 keV-He$^+$-ions at a scattering angle of 135$^o$ with respect to the surface normal [17]. Spin-resolved photoemission experiments are undertaken on beamline Cassiopee at Synchrotron Soleil using 20eV photons impinging upon the sample at 45$^o$ with a horizontally polarized electric field. Photoelectrons are acquired in normal-emission geometry with an energy resolution of 130 meV. Spin contrast is achieved using a Mott detector, which exploits the left-right asymmetry of electron scattering due to spin-orbit interaction [18].

## 3. Results and Discussion

We first discuss the growth of C on fcc Co(001). The morphology and the electrical behavior of the ultrathin C films of varying thickness deposited onto Co were investigated by STM and STS (see Figs. 1 (a-f)).

FIGURE 1

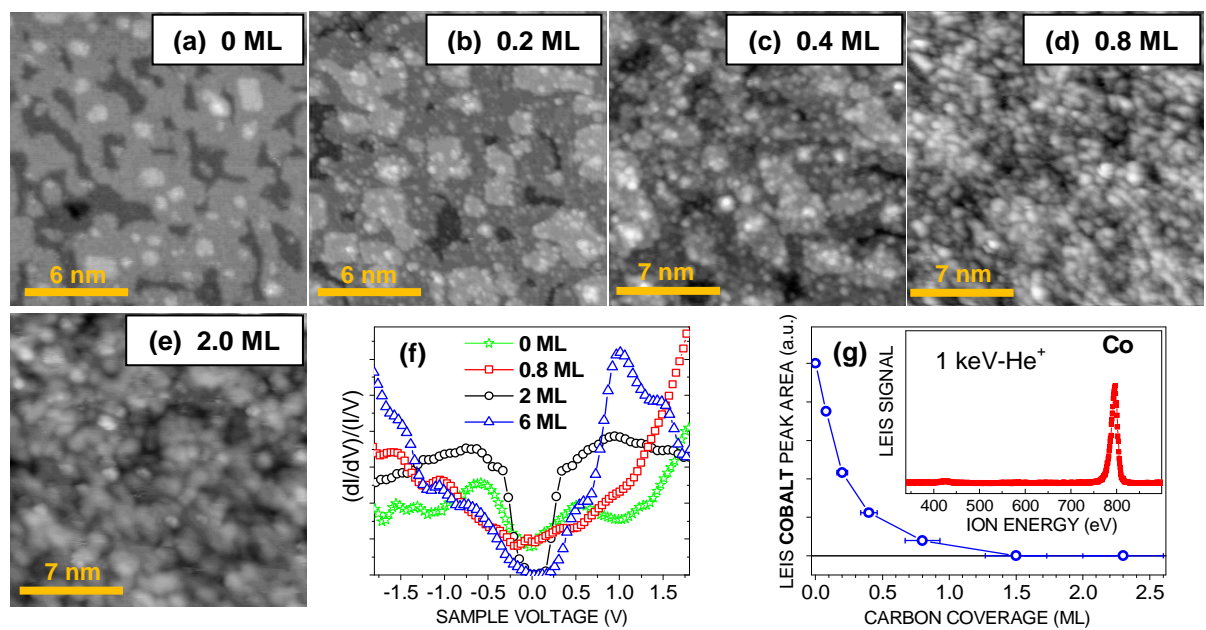

**Figure 1:** (a) STM constant-current images before C deposition and (b-e) for different C coverage: (b) 0.2 ML, (c) 0.4 ML, (d) 0.8 ML and (e) 2 ML. The image size is 40x40 nm$^2$ (a,b) and 50x50 nm$^2$ (c-e). The sample voltages (V) and tunneling current (I) were fixed to +3 V and 0.1 nA, respectively. (f) Normalized tunneling conductance (dI/dV)/(I/V) curves collected before C deposition and for 0.8, 2 and 6 ML C coverage. These spectra have been numerically calculated from I-V curves acquired with the CITS mode (see text; for normalization procedure see Ref. [19]). (g) Evolution of the normalized LEIS Co peak area versus C thickness. Inset: LEIS spectrum collected with 1 keV incident He$^+$ ions before C deposition. The peak located around 795 eV corresponds to Co.

The Co(001) surface is obtained by epitaxial growth of a 3 nm thick film on a Cu(001) single crystal. Previous STM measurements have shown the nearly ideal layer-by-layer growth mode in this system offering atomically flat Co surfaces [20]. For a carbon coverage of 0.2 ML, deposited and measured at room temperature, we observe small, randomly dispersed protrusions on the Co surface. Most of them have an apparent height of 0.06 nm and an apparent diameter of 0.3-0.5 nm, indicating the presence of small C islands. At 0.4 ML coverage, we find a homogeneous distribution of C islands with a mean lateral extension below 1 nm, from which we infer a small diffusion length of C at room temperature. At 0.8 ML, the island size increases significantly. For a C coverage of 2 ML, dense grain-like features entirely cover the Co surface, which becomes difficult to identify (Fig. 1 (e)). This completion of the C overlayer is confirmed by LEIS (see Fig. 1 (g)), STS and XPS measurements (see below). Notably, normalized tunneling conductance curves presented in Fig. 1(f) show that C in the second and additional layers exhibit a small band gap of about 0.4 eV that lies within the 0- few eV range of band gaps for amorphous carbon [21,22]. Furthermore, STS also shows that the Co/C interface itself is metallic.

To understand how the Co/C organic spinterface forms, followed by semiconducting C, we used XPS to track the hybridization state of C. We present in Fig. 2a the C1s core-level peak with increasing C coverage.

FIGURE 2

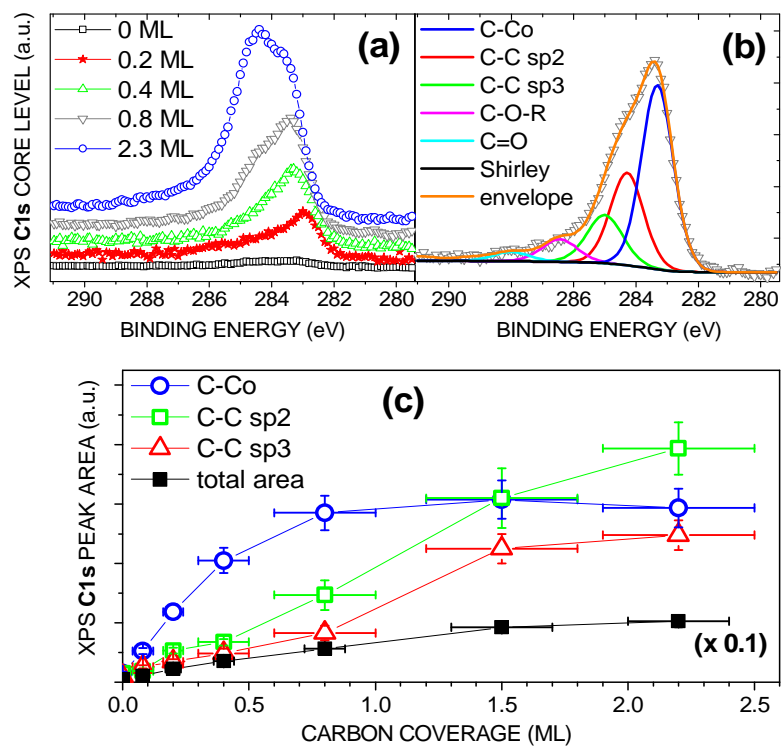

**Figure 2:** (a) C*1s* core-level XPS spectra as a function of C coverage on 20 ML Co/Cu(001). (b) The experimental data (triangles) of a 0.8 ML C film are decomposed into C-Co, C-C $sp^2$, C-C $sp^3$, C-O-R (with R = H or C) and C=O subpeaks after Shirley background subtraction as explained in the text. (c) Area intensity of the C-Co, C-C $sp^2$ and C-C $sp^3$ subpeaks and the experimental C*1s* peak as a function of C coverage.

We find that, for all samples, over 90 % of the C1s peak intensity results from 3 different main components (Fig. 2b): (1) a component with a binding energy of 283.1±0.2 eV that reflects C-Co bonding, i.e. results from carbidic carbon; (2) one at 284.3 eV due to C-C $sp^2$ bonding that results from graphite-like carbon; and (3) one at 285 eV due to C-C $sp^3$ bonding, i.e. diamond-like carbon. Carbon atoms bonded to oxygen (C-O-R, C=O) contribute only weakly to the C1s peak intensity. Although the three main contributions are present for all C coverages investigated the C1*s* peak is dominated by the C-Co component in the submonolayer regime, indicating a strong hybridization between C 2*p* and Co 3*d* electrons at the interface. The latter strongly increases with coverage and saturates for thicknesses above 1 ML (Fig. 2 (c)). The $sp^2$ and $sp^3$ C-C bond contributions increase as well with C coverage, in line with the increase in C island size observed on the STM images. However, they do not saturate above 1 ML C coverage, as expected. For a C coverage above 2 ML both X-ray photoelectron diffraction measurements of the C1s spectra (not shown) and low-energy electron diffraction studies (not shown) do not reveal any diffraction peaks, thus indicating that the carbon layer is amorphous. Our Raman spectroscopy measurements of a 2 ML thick C film on Co(001) also evidence the amorphous character of the carbon film [23]. We thus conclude that, beyond a metallic Co/C spinterface, the semiconducting layer consists of small, densely packed amorphous C clusters with a majority of $sp^2$-bonded carbon atoms and a $sp^2$-to-$sp^3$ ratio between 1.4 and 1.8.

We now describe the spin-polarized properties at and beyond the Co/C interface. Panels (a) and (b) of Fig. 3 respectively show raw majority-spin and minority-spin photoemission spectra for bare Co and C/Co.

FIGURE 3

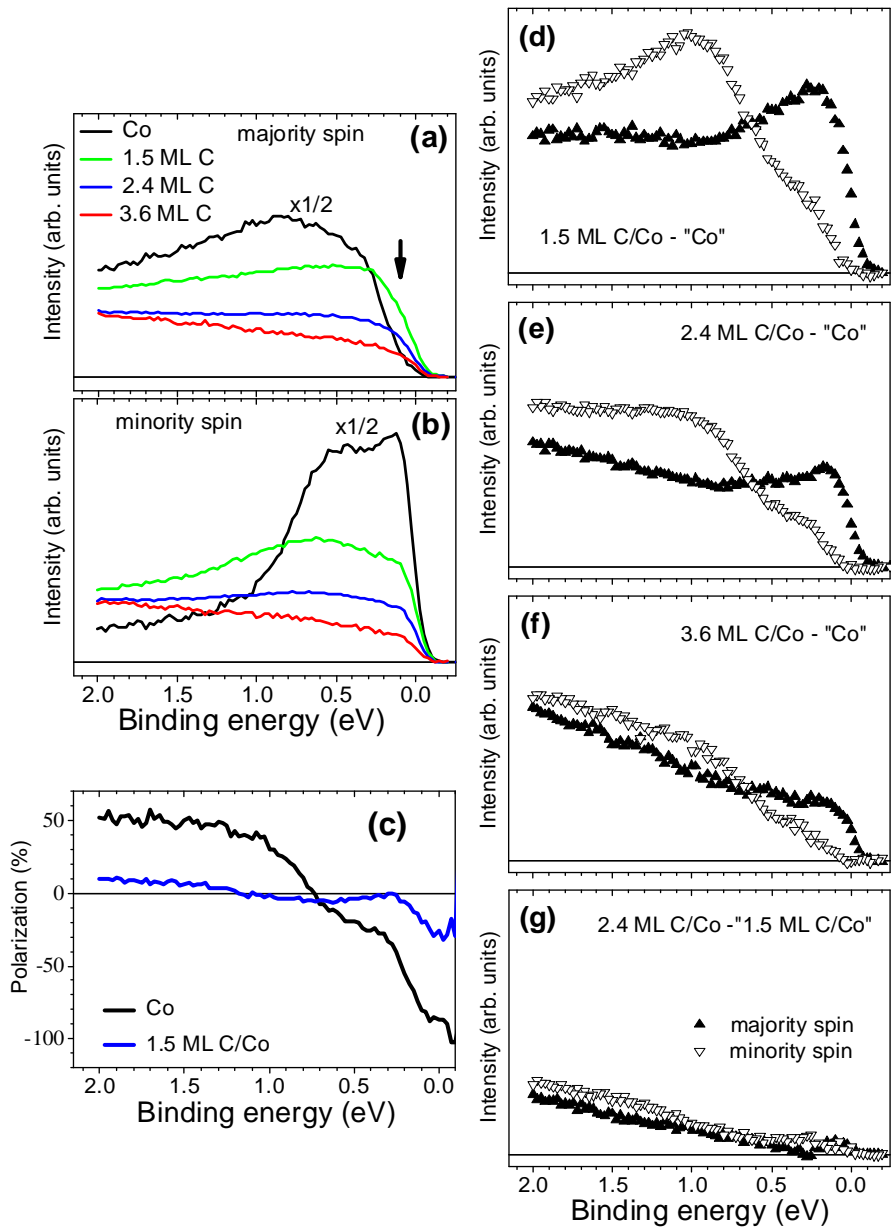

**Figure 3:** (a,b) Spin-resolved photoemission intensity curves as a function of the binding energy for different C coverages. The spectra for pure Co are multiplied by a factor 0.5. (c) Spin polarization as a function of the binding energy for uncovered Co and 1.5 ML C/Co. The arrow in (a) indicates the energy position of an additional C-induced feature in the majority-spin channel. (d-f) The carbon-induced, spin-resolved photoemission intensity as a function of the binding energy after subtraction of a suitably normalized pure Co spectrum ("Co") and (g) after subtraction of a suitably normalized 1.5 ML C/Co spectrum ("1.5 ML C/Co").

The absolute intensities can be compared as all measurements are performed with the same incoming photon intensity. C coverage of Co promotes, in the majority-spin channel, the appearance of an additional structure close to the Fermi energy $E_F$ (see arrow in Fig. 3a), which yields an otherwise absent Fermi edge of the *d* valence band. On the other hand, the photoemission intensity around $E_F$ in the minority-spin channel decreases with increasing C thickness in an exponential manner. This suggests that the minority-spin intensity of uncovered Co is simply attenuated by the C layer near $E_F$ and that no C-induced additional feature is present. Separately, an additional large feature in the minority-spin channel for binding energies above 0.5 eV is present at the interface but disappears for larger C thicknesses.

The appearance of these additional C-induced structures in specific spin channels is summarized when comparing the raw spin polarization spectra of bare and C-covered Co (see Fig. 3 (c)). The spin polarization of bare Co is strongly negative close to $E_F$. C coverage strongly modifies the spin polarization of the Co reference, leading to a bump with a vanishing polarization at a binding energy $E_B$ of 0.25 eV. This suggests that, while C coverage may generally attenuate the negative spin polarization of Co, C coverage also contributes a positive spin polarization due to a feature at $E_B$ = 0.25 eV. Similarly at around $E_B$ = 0.9 eV, a C-induced feature reverses the sign of the spin polarization of the Co reference. To extract the spin-resolved photoemission signal induced only by the presence of the C layer, we adopt a subtraction procedure that takes into account the C-induced attenuation of the Co photoemission signal (see SI and Ref. [12]). Fig. 3 (d) thus shows the photoemission signal arising from solely 1.5 ML C when deposited on Co. Two highly spin-polarized states appear at $E_B$ = 0.25 eV and $E_B$ = 1 eV. Notably, we find that 1.5 ML C exhibits nearly total positive spin polarization around $E_F$ when deposited onto Co.

The photoemission signal coming from these electronic states weakens for thicker C films (see Fig. 3 (e) and (f)), implying that these are interface states whose signal is attenuated by additional C coverage beyond the interface. As confirmation of this point, additional C coverage beyond the interface no longer contributes to the structures at $E_B = 0.25$ eV and $E_B = 1$ eV (see Fig. 3 (g)). Indeed, the spectra in both spin channels are featureless and exhibit only an overall smooth increase with increasing binding energy.

High-efficiency, Co-based spinterfaces can be obtained using Pc molecules that do not form carbide (Co-C) bonds (XPS data not shown) as seen for Co/C interfaces [12]. Furthermore, these bonds are not expected to promote states near $E_F$. We therefore surmise that these bonds, although dominant (see Fig. 2c), do not contribute to the high spin-polarization at $E_F$, but rather $sp^2$–bonds, which are present in both systems.

## 4. Conclusion

In conclusion, our results demonstrate highly spin-polarized interfaces at room temperature for spintronics using mere C atop the simple ferromagnet Co. While similar to results using Pc molecules [12], this shows how highly efficient organic spinterfaces at room temperature constitute a generic effect that isn't specific to a particular molecule.

Looking forward, this newly discovered carbon-based spinterface combines several applicative advantages. First, both the interfacial C layer and ensuing C monolayers are dense. By involving a maximum number of Co sites, this confers maximum robustness to the Co-induced spinterface. Also, by increasing the effective density of the top portion of an organic layer, C could, as with LiF in organic electronics [24], prevent metallic interdiffusion due to counter electrode deposition. This should strengthen the field of organic spintronics by enabling more systematic [7,8] studies of spin-polarized transport over nominally thinner organic layers, toward a better understanding of the mechanism of spin conservation during transport [4,5]. Finally, the low band gap and its tunability in amorphous C [21,22] could prove interesting toward low-resistance nanoscale spintronic devices [25] that integrate C-based spinterfaces.

**Acknowledgements**

We thank F. Ibrahim and M. Alouani for useful discussions. We gratefully acknowledge support from the CNRS, the Institut Carnot MICA's 'Spinterface' grant, from ANR grant ANR-11-LABX-0058 NIE and from the Franco-German university. We thank the SOLEIL staff for technical assistance and insightful discussions.

$^\$$present address: Université de Ferhat Abbas Sétif 1, Faculté de Technologie, Sétif, Algeria